\documentclass[aps,prl,twocolumn,superscriptaddress,showkeys,preprintnumbers]{revtex4-1}

\usepackage{amsmath,amssymb,ifpdf,url,mathrsfs,slashed,multirow,tabularx,type1cm}
\usepackage{subfigure}
\usepackage{graphicx,color}

\ifpdf        
\usepackage{graphics}
\else                          
\usepackage[dvipdfmx]{graphics}      
\fi

\allowdisplaybreaks

\definecolor{BlueViolet}{rgb}{0.2, 0.00, 0.7}
\definecolor{Blue}{rgb}{0.15, 0.00, 0.9}
\usepackage[%dvipdfmx,
colorlinks=true,linkcolor=Blue,citecolor=Blue,
urlcolor=BlueViolet,breaklinks=true]{hyperref}

\newcommand{\GeV}{\,{\rm GeV}}
\newcommand{\MeV}{\,{\rm MeV}}

\newcommand{\Slash}[1]{{\ooalign{\hfil \hspace*{-5pt}~#1\hfil\crcr\raise.167ex\hbox{/}}}}

\def\({\left(}
\def\){\right)}
\def\<{\langle}
\def\>{\rangle}

\newcommand{\matl}{\left( \begin{array}}
\newcommand{\matr}{\end{array} \right)}

\def\beq#1\eeq{\begin{align}#1\end{align}}

\RequirePackage{xspace}

\RequirePackage{xspace}
\def\Bbar    {\kern 0.18em\overline{\kern -0.18em B}{}\xspace}
\def\Bb      {\ensuremath{\Bbar}\xspace}
\def\nub        {\ensuremath{\overline{\nu}}\xspace}

\begin{document}

\preprint{TTP18--013}

\title{
\boldmath 
Soft-Photon Corrections to
 $\Bb \to D \tau^{-}  \nub_{\tau}$ Relative to  $\Bb \to D \mu^{-}  \nub_{\mu}$ 
\unboldmath
}
 \author{Stefan de Boer}
\email{stefan.boer@kit.edu}
\affiliation{Institute for Theoretical Particle Physics (TTP), Karlsruhe Institute of Technology, Engesserstra{\ss}e 7, D-76128 Karlsruhe, Germany}

\author{Teppei Kitahara} \email{teppei.kitahara@kit.edu} 
\affiliation{Institute for Theoretical Particle Physics (TTP), Karlsruhe Institute of Technology, Engesserstra{\ss}e 7, D-76128 Karlsruhe, Germany}
\affiliation{Institute for Nuclear Physics (IKP), Karlsruhe Institute of
Technology, Hermann-von-Helmholtz-Platz 1, D-76344
Eggenstein-Leopoldshafen, Germany}

\author{Ivan Ni\v{s}and\v{z}i\'c}
\email{ivan.nisandzic@kit.edu}
\affiliation{Institute for Theoretical Particle Physics (TTP), Karlsruhe Institute of Technology, Engesserstra{\ss}e 7, D-76128 Karlsruhe, Germany}
  
\date{\today}

\begin{abstract}
We evaluate long-distance electromagnetic (QED) contributions to
$\Bb^0 \to D^+ \tau^{-}  \nub_{\tau}$ and $B^- \to D^0 \tau^{-}  \nub_{\tau}$
relative to 
$\Bb^0 \to D^+ \mu^{-}  \nub_{\mu}$ and $B^- \to D^0 \mu^{-}  \nub_{\mu}$, respectively,
 in the standard model.
We point out that the QED corrections to the ratios $R(D^{+})$ and $R(D^{0})$ are not negligible, contrary to the expectation that radiative corrections are almost canceled out in the ratio of the two branching fractions.
The reason is that long-distance QED corrections depend on the masses and relative velocities of the daughter particles. 
We find that theoretical predictions for $R(D^{+})^{\tau/\mu}$ and $R(D^{0})^{\tau/\mu}$ can be amplified by $\sim4\%$ and $\sim3\%$, respectively, for the soft-photon energy cut in the range $20$--$40$ MeV.
\end{abstract}

\maketitle

%%%%%%%%%%%%%%%%%%%%%%%%%%%%%%%%%%%%%%%%%%%%%
%\section{Introduction}
%%%%%%%%%%%%%%%%%%%%%%%%%%%%%%%%%%%%%%%%%%%%%

The semileptonic $B$ meson decays that are at the elementary level induced by the $b \to c \ell \nub_{\ell}$ transitions provide a potentially interesting avenue for testing the standard model (SM) at low energies. 
In this respect, it turns out useful to construct the ratios $R(H)$, $H=D,D^\ast$ between the branching fractions that involve $\tau$ leptons and those involving light leptons.
These observables do not depend on the Cabibbo--Kobayashi--Maskawa matrix element $V_{cb}$ and are also theoretically cleaner due to the (partial) cancellation of the hadronic uncertainties parametrized by the corresponding form factors. 
The forthcoming Belle\,II experiment is expected to reduce the corresponding measurement uncertainties to the level of around $3\%$ \cite{Golob}, 
comparable to the current theoretical uncertainties.
This is also the typical size of electromagnetic (QED) effects which we turn to study in this Letter, focusing on long-distance QED effects in $R(D)$.

Short-distance electroweak (EW) contributions to branching fractions of semileptonic decays were evaluated to $1.3\% $  \cite{Aoki:2016frl, Sirlin:1981ie, Atwood:1989em}, 
but since such corrections are lepton universal they cancel in the ratio $R(D)$.
The complete understanding of QED effects in meson decays is a nontrivial task due to the complicated interplay with QCD dynamics, {\it e.g.}, structure-dependent contributions that probe the hadronic content \cite{Becirevic:2009fy,Bernlochner:2010yd,Bernlochner:2010fc}.
In this Letter, we evaluate the lepton-mass-dependent soft-photon effects, which give rise to important corrections.

We point out terms that distinguish the cases of the neutral and charged  $B$ decays
\beq
R(D^+) &\equiv \frac{ \mathcal{B}\left( \Bb^0 \to D^+ \tau^{-} \nub_{\tau} \right)}{\mathcal{B}\left( \Bb^0 \to D^+ \ell^{-} \nub_{\ell}  \right)},\\
R(D^0) &\equiv \frac{ \mathcal{B}\left( B^- \to D^0 \tau^{-} \nub_{\tau} \right)}{\mathcal{B}\left( B^- \to D^0 \ell^{-} \nub_{\ell}  \right)}.
\eeq
The up-to-date average \cite{Aoki:2016frl} of the lattice-QCD predictions \cite{Lattice:2015rga, Na:2015kha} is
\beq
R(D^{+})_{\textrm{SM}} = R(D^{0})_{\textrm{SM}} = 0.300 \pm 0.008\,,
\label{Eq:RDSM}
\eeq
which is consistent with previous evaluations involving different approaches (see \cite{Kamenik:2008tj, Becirevic:2012jf, Bigi:2016mdz, Bernlochner:2017jka, Jaiswal:2017rve}).
The corresponding experimental average \cite{Amhis:2016xyh} of the BaBar \cite{Lees:2012xj,Lees:2013uzd} and Belle \cite{Huschle:2015rga} measurements is 
\beq
R(D)_{\textrm{exp}} = 0.403\pm0.040\pm0.024 \,,
\eeq
which combines electrons and muons for the decay into the light lepton and averages neutral and  charged $B$ decays.
The averaged experimental result exceeds the SM expectation at the level of $2.2\sigma$.
Combined with current discrepancy with respect to the SM in $R(D^{\ast})$, these have been considered as a hint of physics beyond the SM.

One should note that 
the measured results partially include soft photons using \textsc{PHOTOS} 
Monte-Carlo generator \cite{Belle1, Belle2}
for the simulation of modifications of the kinematic variables induced by final-state photon radiations 
 \cite{Barberio:1990ms,Barberio:1993qi,Davidson:2010ew}. 
To our knowledge, our results are not fully covered by \textsc{PHOTOS} for $\Bb \to D \ell \nub_{\ell}$ \cite{footnote1}; {\it e.g.}, 
we include interferences between initial- and final-state emissions and virtual corrections including the Coulomb terms.

For previous studies of QED effects in (semi)leptonic $B$ decays, we refer the reader to Refs.~\cite{Cirigliano:2005ms,Becirevic:2009aq,Bernlochner:2010yd,Bernlochner:2010fc,Tostado:2015tna}.
Related works regarding $b\to s\ell^+\ell^-$ transitions can be found in Refs.~\cite{Buras:2012ru, Bordone:2016gaq,Beneke:2017vpq}.

\begin{figure}[t]
\begin{center} 
\includegraphics[width=0.4 \textwidth,
bb  = 0 0 919 340
]{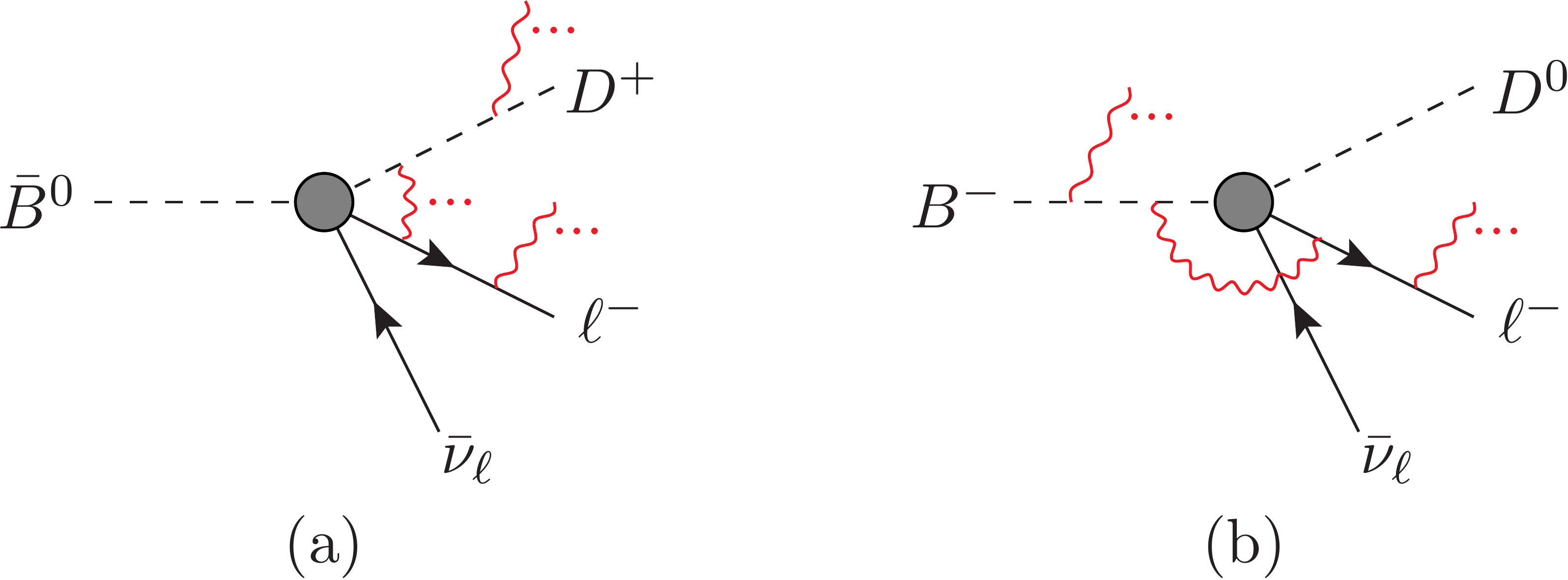}
\vspace{-0.2cm}
\caption{Soft-photon contributions to (a) $R(D^+)$ and (b) $R(D^0)$, where the self-energy diagrams 
and loop diagrams induced by $\Bb D{\ell}\bar{\nu}\gamma$ vertex are omitted for simplicity. 
The dots represent an arbitrary number of soft photons. 
}
\label{fig:diagram}
\end{center}
\vspace{-0.8cm}
\end{figure}

\vspace{-0.2cm}
%%%%%%%%%%%%%%%%%%%%%%%%%%%%%%%%%%%%%%%%%%%%%%%%%%%%%%%%%%%%
\boldmath  
\section{
QED corrections in $\Bb \to D \ell \nub_\ell$}\unboldmath
%%%%%%%%%%%%%%%%%%%%%%%%%%%%%%%%%%%%%%%%%%%%%%%%%%%%%%%%%%%%
\vspace{-0.15cm}

In this section, we calculate the QED corrections to the processes $\Bb \to D\ell\nub_{\ell}$ ($\ell=\mu,\tau$) at large distances, where the electromagnetic interactions of the charged scalar mesons are well described by the scalar QED.

The correction factors exhibit dependence on the kinematic variables $s_{D \ell} \equiv \left( p_{D} + p_{\ell} \right)^2$ and $q^2 \equiv \left( p_B - p_{D} \right)^2 = \left( p_{\ell} + p_{\nu} \right)^2$, which require the double differential decay distribution.
At the tree level, it reads

\beq
&\frac{ d^2 \Gamma_0}{d q^2 d s_{D\ell}}
= \frac{ G_F^2 \left|V_{cb} \right|^2}{512 \pi^3 m_B^3}\eta_\textrm{EW}^2  \Bigl\{ f_0 (q^2) f_{+} (q^2) a_{0+} (q^2, s_{D \ell})  \nonumber \\
&~~~~~~~~+  \left[ f_{+} (q^2)\right]^2  a_{+}(q^2, s_{D \ell})  +   
\left[ f_0 (q^2)\right]^2 a_{0}(q^2)   \Bigl\}  , 
\label{eq:tree}
\eeq
including also the short-distance corrections $\eta_\textrm{EW} = 1.0066$ \cite{Aoki:2016frl, Sirlin:1981ie, Atwood:1989em}, and the coefficients $a_{0+},\,a_{+}$, and $a_{0}$ are given by
\begin{widetext}
\vspace{-0.5cm}
\beq
a_{0+}  =& 8 (q^{2})^{-2} m_{\ell}^2 ( m_B^2 - m_D^2)  \left[   (m_D^2 - q^2)(q^2 - m_{\ell}^2) + m_B^2 ( q^2 + m_{\ell}^2) - 2 q^2 s_{D \ell} \right]\,,\\
a_{+} = &4 (q^{2})^{-2} \left\{     m_{\ell}^2 (m_D^2 -q^2)^2 (q^2  - m_{\ell}^2) - m_B^4 (m_{\ell}^4 + 3 m_{\ell}^2 q^2) + 4 q^2 s_{D \ell} (m_{\ell}^2 - q^2 )  ( q^2 - m_D^2)   \right. \nonumber \\
& ~~~~~~~~~~~\left.  - 4 (q^2)^2 s_{D \ell}^2 + 2 m_B^2 \left[ ( m_{\ell}^2 -q^2) \left[- m_{\ell}^2 q^2 + m_D^2 (m_{\ell}^2 + 2 q^2) \right]+ 2 q^2 s_{D \ell} (m_{\ell}^2 + q^2) \right]\right\},\\
a_{0}  =& 4 (q^{2})^{-2} m_{\ell}^2 (m_B^2 - m_D^2)^2  ( q^2- m_{\ell}^2)\,.
\eeq
\vspace{-0.5cm}
\end{widetext}
The corresponding boundaries of the phase space integral can be found in Ref.~\cite{Patrignani:2016xqp}. 
For the form factors $f_0(q^2)$ and $f_{+}(q^2)$ we use the averaged results from Ref.~\cite{Aoki:2016frl}.

Adding the long-distance QED contributions from real photon emissions and virtual corrections, we obtain the following compact formulas for the decay process $\Bb^0 \to D^+ \ell^- \nub_{\ell}$ (see Fig.~\ref{fig:diagram}), where $\ell = \mu,\,\tau$,
\beq
\frac{ d^2 \Gamma}{d q^2 d s_{D\ell}}  =&   \frac{ d^2 \Gamma_0}{d q^2 d s_{D\ell}} \Omega_B^{D^+} \Omega_C   \bigl[    1 +  
\frac{\alpha}{\pi} \bigl(   F_{D} + F_{\ell}  \nonumber \\
&  
  -2  F_{D \ell}  -2  H_{D \ell} 
\bigr) \bigr] +\frac{\alpha}{\pi}  \frac{ d^2 \tilde{\Gamma}^{D^+}}{d q^2 d s_{D\ell}} \,,\label{eq:d2Gamma_neutralB}
\eeq
with $\alpha = 1/137$, and for $B^- \to D^0 \ell^- \nub_{\ell}$,
\beq
\frac{ d^2 \Gamma}{d q^2 d s_{D\ell}}  &=   \frac{ d^2 \Gamma_0}{d q^2 d s_{D\ell}} \Omega_B^{D^0}  \bigl[    1 +  
\frac{\alpha}{\pi} \bigl(   1 + F_{\ell} \nonumber \\
&  -2  F_{B \ell}  -2  H_{B \ell} 
\bigr) \bigr] +\frac{\alpha}{\pi}  \frac{ d^2 \tilde{\Gamma}^{D^0}}{d q^2 d s_{D\ell}}\,,\label{eq:d2Gamma_chargedB}
\eeq
following the notation from Ref.~\cite{Isidori:2007zt}.
For the derivations of Eqs.~(\ref{eq:d2Gamma_neutralB}) and (\ref{eq:d2Gamma_chargedB}) we adopt the soft-photon approximation \cite{Jauch,Yennie:1961ad, Weinberg:1965nx}, including terms $\mathcal O(\ln E_\textrm{max})$ and $\mathcal O(E_\textrm{max}^0)$ \cite{Isidori:2007zt}, where $E_{\textrm{max}}$ is the maximum total energy  of undetected soft photons in the rest frame of the $\Bb$ meson.
We analytically checked that the infrared (IR) divergences cancel.
We describe each of the terms appearing in Eqs.~(\ref{eq:d2Gamma_neutralB}) and (\ref{eq:d2Gamma_chargedB}) separately in the following.

The only coefficients that depend on $E_{\textrm{max}}$ are
\beq
\Omega_B^{D^+} & = \left( \frac{ 2 E_{\textrm{max}}}{ \sqrt{ m_D m_{\ell}}} \right)^{- \frac{ 2 \alpha}{\pi}( 1 - 2 b_{D \ell} )}\,, 
\label{Eq:omegaB}\\
\Omega_B^{D^0} & =  \left( \frac{ 2 E_{\textrm{max}}}{ \sqrt{ m_B m_{\ell}}} \right)^{- \frac{ 2 \alpha}{\pi}( 1 - 2 b_{B \ell} )}\,,
\label{Eq:omegaB2}
\eeq
where we resum the potentially large contributions $\left( \alpha \ln E_{\textrm{max}} \right)^n $ to all orders (see Fig.~\ref{fig:diagram}), following Refs.~\cite{Isidori:2007zt, Weinberg:1965nx}.
Here, for $i=D,\,B$,
\beq
b_{i  \ell} &  = \frac{1}{4 \beta_{i \ell}} \ln  \frac{ 1 + \beta_{i \ell} }{ 1 - \beta_{i  \ell} } \,,\\
\beta_{D \ell} & = \left[  1 - \frac{ 4 m_D^2 m_{\ell}^2}{\left(s_{D \ell} - m_D^2 - m_{\ell}^2 \right)^2} \right]^{\frac{1}{2}}\,, \\
\beta_{B \ell} & =  \left( 1 - \frac{ m_\ell^2}{E_\ell^2} \right)^{\frac{1}{2}}, \quad E_{\ell} = \frac{s_{D \ell} + q^2 - m_D^2}{2 m_B}\,,\label{eq:El} 
\eeq
where $E_\ell$ is the energy of the charged lepton in the rest frame of the $\Bb$ meson and $\beta_{ij}$ ($0 < \beta_{ij} <1 $) denotes the relative velocity of the particles $i$ and $j$ in the rest frame of either particle.

The Coulomb resummation of the $\left( \pi \alpha / \beta_{D \ell} \right)^n$ terms (Sommerfeld enhancement \cite{Sommerfeld}) is denoted by $\Omega_C$,
which for a fermion-scalar pair is given by \cite{Hryczuk:2011tq}
\beq
\Omega_C = - \frac{ 2 \pi \alpha }{ \beta_{D \ell}} \frac{1}{ e^{- \frac{ 2 \pi \alpha }{ \beta_{D \ell}} } -1  }\,.
\label{Eq:Coulomb}
\eeq
The effect of this resummation with respect to the corresponding leading-order term turns out negligible in the final integrated rates.
We also find that the corresponding Coulomb term is absent in the case of the charged $B$ decay. 

We note that $D^+$ and $\tau^-$ are sufficiently long-lived for the resummations to be valid \cite{Falgari:2012sq} [$\Gamma /m \sim \mathcal{O}(10^{-12 }) \ll  \mathcal{O}(0.1) \lesssim \beta_{D \ell}$].

We checked that expansions of the resummation factors in $\alpha$ agree with explicit calculations of the soft-photon emissions and the virtual corrections.

Finally, the energy-independent terms $F$ represent the real photon emissions, while the terms denoted by
$H$ correspond to virtual corrections without the Coulomb term.
They read, for $i=D,\,\ell$,
\beq
F_{i} & = \frac{1}{2 \beta_{Bi} }\ln  \frac{ 1 + \beta_{Bi} }{ 1 - \beta_{Bi} }  \,,\label{eq:Fi}
\eeq
\begin{widetext}
and, for $ij=D\ell,\,B\ell$,
\beq
F_{D \ell} &  = 
\frac{1}{2} \frac{ m_D m_{\ell} }{ \sqrt{ 1 - \beta_{D \ell}^2  }} 
\int^{1}_{0} 
d z \frac{E(z)}{P(z) 
\left[ E(z)^2 -  P(z)^2 \right]} 
\ln  \frac{ E(z) + P(z)   }{ E(z) - P(z)   }\label{eq:FDL}
\,,\\
F_{B \ell}&  = \frac{1}{4 \beta_{B \ell}} \left\{ \textrm{Li}_2\left(    \frac{ 1 - \beta_{B \ell}}{2} \right) 
- \textrm{Li}_2\left(    \frac{ 1 + \beta_{B \ell}}{2} \right)  + 4 \textrm{Li}_2\left(     \beta_{B \ell} \right)  -  \textrm{Li}_2\left(  \beta_{B \ell}^2 \right)   + \ln2\, \ln \frac{ 1+ \beta_{B \ell}}{1 - \beta_{B \ell}} \right. \nonumber \\
& ~~~~~~~~~~~~~    \left. + \frac{1}{2} \ln^2 \left( 1 - \beta_{B \ell} \right)    - \frac{1}{2} \ln^2 \left( 1 + \beta_{B \ell} \right)    \right\}\,,\label{eq:FBl}\\
H_{ij}
&= 
   -\frac{1}{2 \beta_{ij}} \Biggl\{  \frac{1}{2} \ln^2\frac{m_i}{m_j}- \frac{1}{8} \ln^2\frac{1+\beta_{ij}}{1-\beta_{ij}} - \frac{1}{2} \ln^2\left| \frac{\Delta^i_{ij} +\Delta_{ij} \beta_{ij}}{\Delta^j_{ij} +\Delta_{ij} \beta_{ij}} \right|  - \textrm{Li}_2 \left( \frac{2 \Delta_{ij} \beta_{ij}}{\Delta^i_{ij} +\Delta_{ij} \beta_{ij}} \right) - \textrm{Li}_2 \left( \frac{2 \Delta_{ij} \beta_{ij}}{\Delta^j_{ij} +\Delta_{ij} \beta_{ij}} \right)\Biggr\}\nonumber\\
&~~~~+\frac{1}{4} \ln \frac{m_i m_j  }{\mu^2}    -\frac{1}{2}
- \frac{ m_i^2 - m_j^2}{4 s_{ij}}  \ln \frac{m_i}{m_j}  - \frac{1}{4}   \Delta_{ij} \beta_{ij} \ln \frac{1 + \beta_{ij}}{1 - \beta_{ij}} 
  - \frac{\Delta_{i j }}{2} \ln \frac{m_{i}}{m_{j}} - \frac{\Delta_{i  j}^{i}}{ 4 \beta_{i  j}} 
  \ln \frac{ 1 + \beta_{ij}}{ 1 - \beta_{ij}}
   \,,\label{eq:Hij}
\eeq
\end{widetext}
where
\beq
&\Delta_{ij }  = \frac{ s_{ij } - m_i^2 - m_{j}^2}{2 s_{ij}}\,,\;\Delta^{i,j}_{ij}  = \frac{ s_{ij}  + m_{i,j}^2 - m_{j,i}^2}{2 s_{ij }} \,,\\
&s_{B \ell} \equiv (p_B - p_{\ell})^2 = m_B^2 + m_D^2 + m_{\ell}^2  - q^2  -  s_{D \ell}\,,\\
&\textrm{Li}_2 (z) \equiv   -  \int^{z}_0 d t \frac{ \ln (1 - t ) }{t} \,.
\eeq
The functions $E(z)$ and $P(z)$ in Eq.~(\ref{eq:FDL}) are given by
\beq
E(z) &=  z E_D + \left( 1 - z \right) E_{\ell}\,,\\
P(z) & =  \Biggl\{ \left[ z E_D + \left( 1 - z \right) E_{\ell} \right]^2  - z^2 m_D^2  \nonumber \\
& ~~~~  - \left( 1 - z \right)^2 m_{\ell}^2 - 2 z (1-z) \frac{m_D m_{\ell}}{\sqrt{  1 - \beta_{D \ell}^2  } }  \Biggr\}^{\frac{1}{2}}\,,
\eeq
and $\beta_{BD}$ is obtained from Eq.~(\ref{eq:El}) by replacing $\ell$ by $D$ and using $E_{D} = \left( m_{B}^2 + m_{D}^2 - q^2 \right) / 2 m_B$.

Note that the IR-finite and $E_{\rm max}$-independent terms $\tilde{\Gamma}$  in Eqs.~(\ref{eq:d2Gamma_neutralB}) and (\ref{eq:d2Gamma_chargedB})  represent loop corrections which can not be factorized from the tree-level decay distribution in Eq.~\eqref{eq:tree} including also terms induced by the $\Bb D{\ell}\bar{\nu}\gamma$ vertex \cite{Bernlochner:2010fc,Kubis:2006nh} in the soft-photon approximation.
We include these contributions in our results; however,  
since their numerical effects are small [$\alpha/\pi \times \mathcal{O}(1)$], we will report the lengthy analytical expressions elsewhere.

Using the independence of soft-photon emission terms on the spins of the external legs \cite{Weinberg:1965nx}, we
checked that Eqs.~(\ref{eq:Fi})--(\ref{eq:FBl}) are in agreement with the corresponding terms from the decay process involving scalar particles evaluated in Ref.~\cite{Isidori:2007zt}.

For ultraviolet divergences, we use the $\overline{\textrm{MS}}$ scheme denoting the renormalization scale as $\mu$, while for the charged-particle self-energies, we adopt the on shell renormalization scheme.
We regularize the IR divergences with a spurious photon mass.

For the derivation of Eq.~(\ref{eq:Hij}), we utilize the analytical result for the three-point one-loop scalar integral given in Ref.~\cite{Beenakker:1988jr}.
We cross-checked the resulting analytic formula for $H$ with the numerical evaluations using \textsc{LoopTools} \cite{Hahn:1998yk} and \textsc{Package-}\textbf{X} \cite{Patel:2015tea}.
The first line of Eq.~(\ref{eq:Hij}) arises from soft virtual photons, while the second line involves remaining terms from the full virtual momentum dependence, neglecting the potential modifications of the momentum dependence of the form factors.

\begin{figure}[t]
\begin{center} 
\subfigure[]
{\includegraphics[width=0.23 \textwidth, 
bb  = 0 0 301 321
]{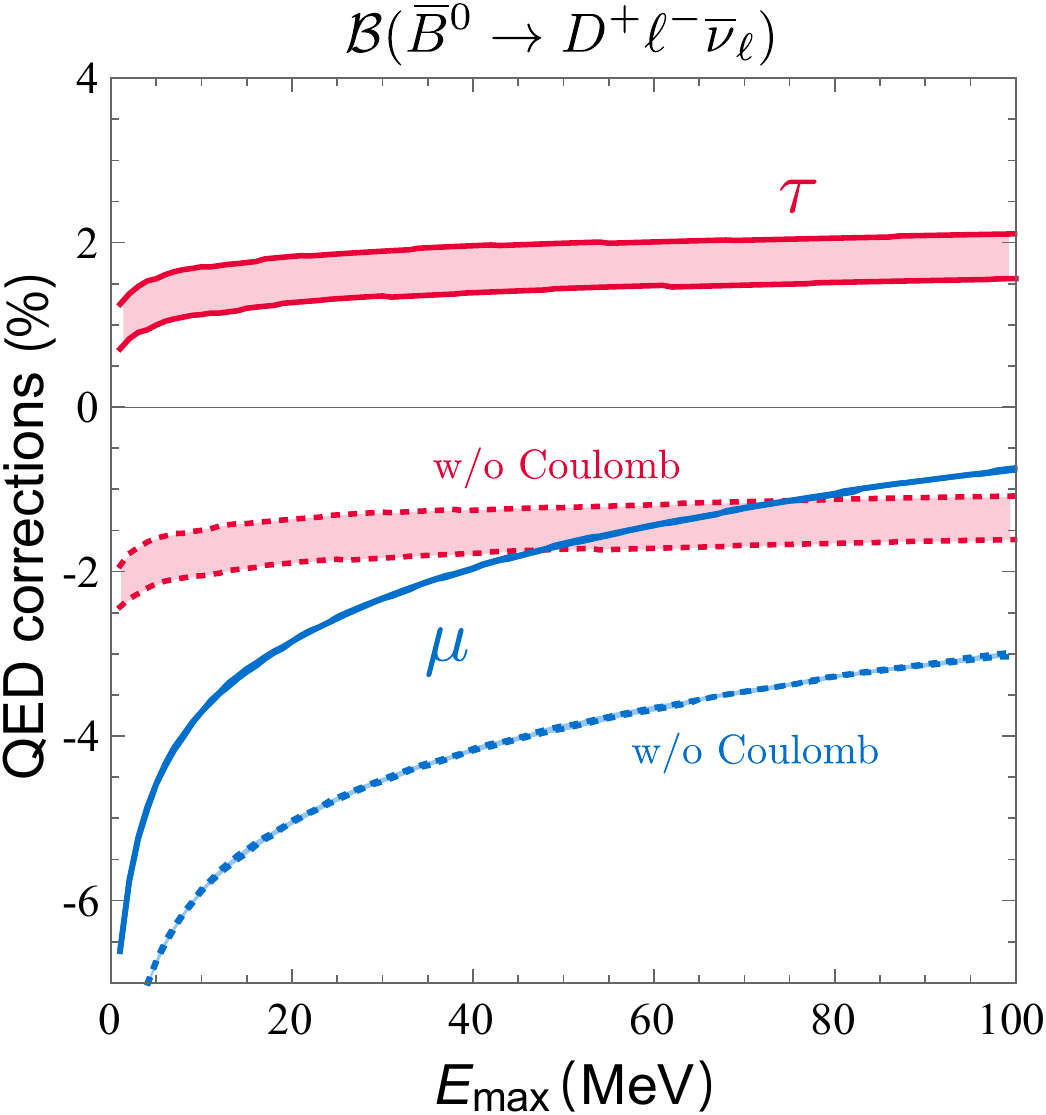}
}\,
\subfigure[]
{\includegraphics[width=0.23 \textwidth, 
bb  = 0 0 301 322
]{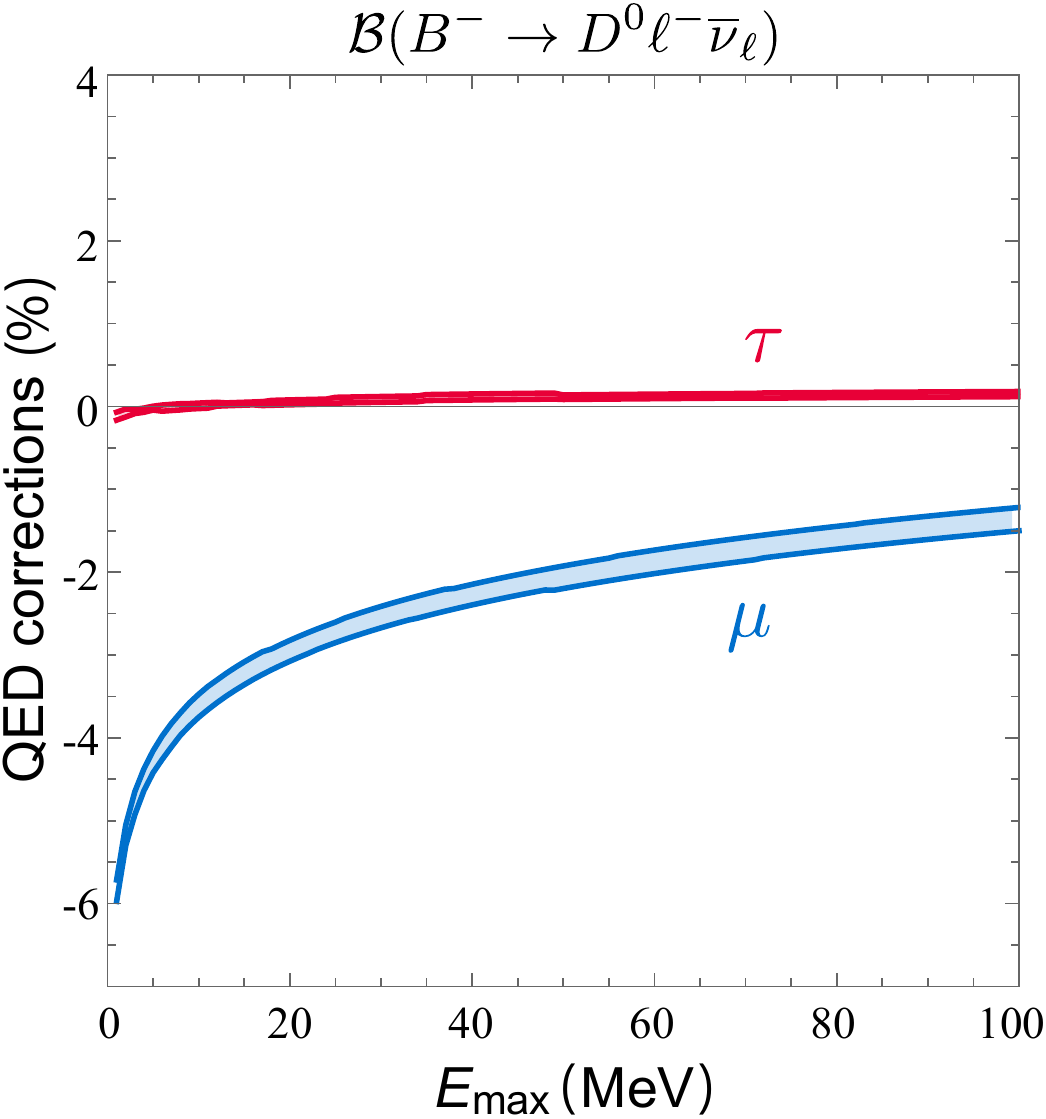}
}
\vspace{-0.2cm}
\caption{ 
The long-distance QED corrections to the branching ratios of (a) $\Bb^0 \to D^+ \ell^- \nub_{\ell}$ and (b) $B^- \to D^0 \ell^- \nub_{\ell}$, where $\ell = \mu,\,\tau$, as a function of $E_{\textrm{max}}$. 
The dotted lines show the corrections to $\Bb^0 \to D^+ \ell^- \nub_{\ell}$ without (w/o) the Coulomb contributions, for the purpose of illustration.
}
\label{fig:Emax_dependent}
\end{center}
\vspace{-0.6cm}
\end{figure}

We refrain from applying the soft-photon approximation to the case of the electron mode, because $m_e \ll E_{\textrm{max}}$ leads to an additional large (Sudakov) logarithm and large finite terms $\mathcal O(E_\textrm{max}/m_e)$, which break the underlying assumption of the approximation (see Ref.~\cite{Kubis:2010mp}). 
We hope to revisit this issue in a future work.

%\vspace{-0.4cm}
%%%%%%%%%%%%%%%%%%%%%%%%%%%%%%%%%%%%%%%%%%%%%%%%%%%%%%%%%%%%
\boldmath  
\section{
Numerical result: $E_{\textrm{max}}$ dependence}\unboldmath
% %%%%%%%%%%%%%%%%%%%%%%%%%%%%%%%%%%%%%%%%%%%%%%%%%%%%%%%%%%%%

In Fig.~\ref{fig:Emax_dependent}, we show the results for the long-distance QED corrections to 
$\mathcal{B}(\Bb^0 \to D^+ \ell^- \nub_{\ell})$ (left panel) and $\mathcal{B}(B^- \to D^0 \ell^- \nub_{\ell})$  (right panel), where $\ell = \mu,\,\tau$, as a function of $E_\textrm{max}$.
Note that the typical value of $E_{\textrm{max}}$ in current experiments is $20$--$30\,\MeV$.
The bands correspond to $100\,\text{MeV}<\mu<1\,\text{GeV}$, where the $\mu$ dependence turns out to be negligible for $\mathcal{B}(\Bb^0 \to D^+ \mu^- \nub_{\mu})$.
To illustrate the impact of the Coulomb contributions, we also show $\mathcal{B}(\Bb^0 \to D^+ \ell^- \nub_{\ell})$ with $\Omega_C$ set to 1 in Eq.~(\ref{eq:d2Gamma_neutralB}).

\begin{figure}[t]
\begin{center} 
\includegraphics[width=0.4 \textwidth, 
bb=0 0 360 239
]{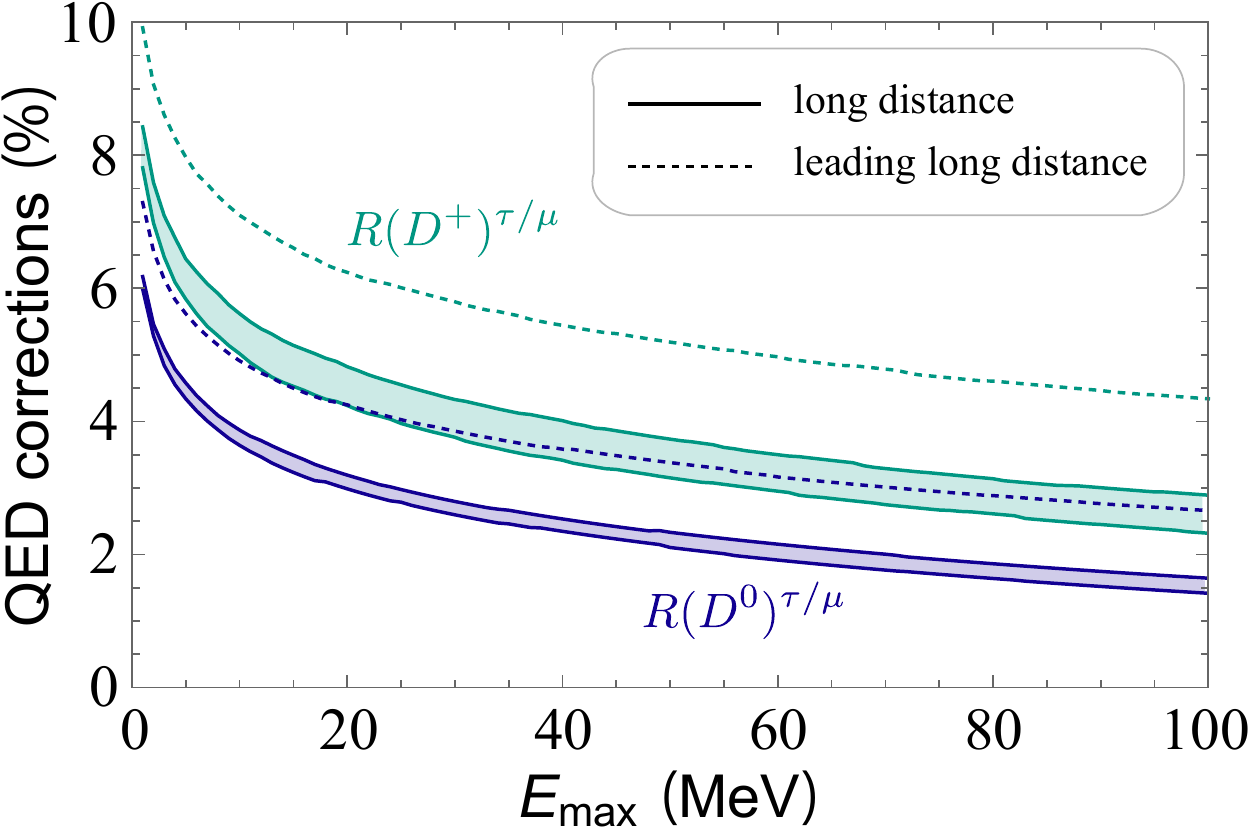}
\vspace{-0.2cm}
\caption{ 
The (leading) long-distance QED corrections to $R(D^{+})^{\tau/\mu}$  and $R(D^{0})^{\tau/\mu}$ as a function of $E_{\textrm{max}}$.
}
\label{fig:Emax_mu_dependent}
\end{center}
\vspace{-0.6cm}
\end{figure}

We observe that the corrections to $\tau$ modes are almost independent of $E_{\textrm{max}}$. 
This can be understood in the nonrelativistic region of Eqs.~\eqref{Eq:omegaB} and \eqref{Eq:omegaB2}, where
\beq
\left( 2 E_{\textrm{max}} \right)^{- \frac{2 \alpha}{\pi} (1 - 2 b_{ij})} \simeq 1+  \frac{2 \alpha}{3 \pi}\ln\left( 2 E_{\textrm{max}} \right)\beta_{ij}^2\,;
\eeq
hence, the $E_{\textrm{max}}$ dependence is suppressed by the small relative velocity involving $\tau$ leptons. 
On the other hand, the corrections to $\mu$ modes are sensitive to $E_{\textrm{max}}$ and negative.
The total effects to the ratios $R(D^{+})$ and $R(D^{0})$ are, therefore, positive and dependent on $E_{\textrm{max}}$ from the muonic modes.
Furthermore, one observes that the Coulomb contribution to the $\tau$ mode is larger than the one to the $\mu$ mode because of the smaller relative velocity in the former case.

Figure~\ref{fig:Emax_mu_dependent} is our main result.
We show the long-distance QED corrections to $R(D^{+})^{\tau/\mu}$ and $R(D^{0})^{\tau/\mu}$, where we define them as the ratios of $\tau$ and $\mu$ modes
and use the same $E_{\textrm{max}}$ for both type of leptons.
Again, the bands correspond to $100\,\text{MeV}<\mu<1\,\text{GeV}$.
We observe that the corrections to $R(D^{+})^{\tau/\mu}$ and $R(D^{0})^{\tau/\mu}$ differ by 0.7\%--1.8\% and propose to properly weight charged and neutral decays in averaging $R(D)$.
We find that the individual corrections are comparable to or larger than the current uncertainty of $R(D)_{\textrm{SM}}$ given in Eq.~\eqref{Eq:RDSM}.
Choosing $E_{\textrm{max}} = 20\,\MeV$ and $\mu = 200$\,\MeV, 
$R(D^{+})_{\textrm{SM}}^{\tau/\mu}$ and $R(D^{0})_{\textrm{SM}}^{\tau/\mu}$ can be amplified by 4.4\% and 3.1\%, respectively.
 We find that the dominant renormalization scale dependence comes from $\tilde{\Gamma}$  in Eqs.~(\ref{eq:d2Gamma_neutralB}) and (\ref{eq:d2Gamma_chargedB}). 
To estimate the potential impacts by the modifications of the momentum dependence of the form factors from virtual loop momenta, we compare our full ({\it long distance}) results to the ones ({\it leading long distance}) that discard the second line in Eq.~(\ref{eq:Hij}) and $\tilde{\Gamma}$.
We obtain a difference of $\sim1.5$--$2\%$ in $R(D^{+})^{\tau/\mu}$ and $\sim1\%$ in $R(D^{0})^{\tau/\mu}$, which indicates that the impacts are subleading.

%\vspace{-0.4cm}
%%%%%%%%%%%%%%%%%%%%%%%%%%%%%%%%%%%%%%%%%%%%%%%%%%%%%%%%%%%%
\boldmath  
\section{
Numerical result: $M_{\textrm{miss}}^2$ dependence}\unboldmath
%%%%%%%%%%%%%%%%%%%%%%%%%%%%%%%%%%%%%%%%%%%%%%%%%%%%%%%%%%%%

In order to relate our formulas to experimental analyses that fit the missing mass squared ($M_\textrm{miss}^2$) distribution, we consider long-distance QED corrections as a function of
\beq
M_{\textrm{miss}}^2 &\equiv \left( p_{e^+ e^-} - p_{B_{\textrm{tag}}} - p_{D} - p_{\ell} \right)^2,
\eeq
where 
$p_{e^+ e^-}$, $p_{B_{\textrm{tag}}}$, $p_{D}$, and $p_{\ell}$ are the four-momenta of the $e^+ e^-$ beams, tagged $B$, and signal $\Bb$ daughter particles, respectively.
The distribution is dominated 
by the detector resolution of these four-momenta, giving a symmetric shape \cite{Belle2}.
We estimate the single soft-photon contribution as
\beq
 M_{\textrm{miss},\gamma}^2  = \left( p_{\nu} + p_{\gamma} \right)^2 = 2 E_{\nu} E_{\gamma} \left( 1 - \cos \theta_{\nu \gamma} \right)>0 \,,
\eeq
where $\theta_{\nu \gamma} $ is the angle between $\nub_{\ell}$ 
and the soft photon.
Hence, single soft photons give only positive contributions to the missing mass squared, resulting in an asymmetric  distribution.
Assuming an isotropic distribution for $\theta_{\nu \gamma} $ gives $M_{\textrm{miss},\gamma}^2\approx 2 E_{\nu} E_{\gamma}$.
Using $E_{\nu}  = \left( m_{B}^2 - s_{D \ell} \right) / 2 m_B$,
we estimate the soft-photon energy as
\beq
E_{\gamma} \lesssim E_{\textrm{max}} \approx \frac{ m_B}{ m_B^2 - s_{D \ell}} \hat M_{\textrm{miss},\gamma}^2\,,\label{eq:Emax_Mmiss}
\eeq
where $\hat M_{\textrm{miss},\gamma}^2$ corresponds to the maximal missing mass squared from single photon emissions.
For instance, 
using  $\hat M_{\textrm{miss},\gamma}^2=0.1\GeV^2$ and $s_{D \ell} = 10 \GeV^2$, one obtains  $E_{\textrm{max}} \approx 30\MeV$.

Substituting Eq.~(\ref{eq:Emax_Mmiss}) into Eqs.~(\ref{Eq:omegaB}) and (\ref{Eq:omegaB2}), we assess the long-distance QED corrections to 
$\mathcal{B}(\Bb^0 \to D^+ \mu^- \nub_{\mu})$ as 
$\{-2.8, -1.9, -1.0\}\%$ and to $\mathcal{B}(B^- \to D^0 \mu^- \nub_{\mu})$ as 
$\{-2.9, -2.3, -1.6\}\%$ for $\hat M_{\textrm{miss},\gamma}^2=\{0.05,0.1,0.2\}\,\text{GeV}^2$, respectively, at $\mu  = 200$\,\MeV.
Note that the above analysis can not be applied for the $\tau$ lepton because of additional neutrinos from its subsequent decay; however, the $\tau$ mode is insensitive to $E_\textrm{max}$ (see Fig.~\ref{fig:Emax_dependent}).

%\vspace{-0.4cm}
%%%%%%%%%%%%%%%%%%%%%%%%%%%%%%%%%%%%%%%%%%%%%
\section{Conclusions}
%%%%%%%%%%%%%%%%%%%%%%%%%%%%%%%%%%%%%%%%%%%%%

We evaluate the soft-photon corrections to $R(D^+)^{\tau/\mu}$ and $R(D^0)^{\tau/\mu}$ as a function of the photon energy cut, see Fig.~\ref{fig:Emax_mu_dependent}.
For example, by taking $E_{\textrm{max}} = 20\,\MeV$, we find that $R(D^{+})^{\tau/\mu}_{\textrm{SM}}$ and $R(D^{0})^{\tau/\mu}_{\textrm{SM}}$ can be amplified by 4.4\% and 3.1\%,  respectively, which are larger than the current lattice-QCD uncertainty of $R(D)_{\textrm{SM}}$.
We emphasize the impact of lepton-mass-dependent contributions and to distinguish between neutral and charged $B$ decays.
Note, however, that a caution is required for introducing the presented effects into the comparisons of the theoretical observables and the available measurements for two reasons:
the effects depend on the precise details of the measurements regarding the cuts related to photon emissions and also involve the electron modes for which we presently do not evaluate a prediction.
We would also like to reiterate that our analysis is valid in the soft-photon region only in which the cut on the photon energy is small relatively to other mass scales in the problem.
Evaluations of the totally photon-inclusive rates would require nonperturbative treatments, for which one could adopt some models, {\it e.g.}, effect of the intermediate excited $D$ resonances \cite{Becirevic:2009fy} and/or modifications of the $q^2$ dependence of the form factors due to the momenta transfer by the hard photons \cite{Bernlochner:2010yd,Bernlochner:2010fc}.
Analogous calculations could also be performed for the case of $R(D^*)$ but are beyond the scope of this Letter.
We expect that the careful treatment of the electromagnetic effects is going to be important for the analyses of future precise measurements.

\vspace{-.4cm}
%%%%%%%%%%%%%%%%%%%%%%%%%%%%%%%%%%%%%%%%%%%%%
\section*{Acknowledgments}
%%%%%%%%%%%%%%%%%%%%%%%%%%%%%%%%%%%%%%%%%%%%%

We would like to thank Florian Bernlochner, Pablo Goldenzweig, and Shigeki Hirose for many kinds of  valuable comments about the Belle experiment.
We are also grateful to 
Monika Blanke,
Thomas Deppisch,
Motoi Endo, 
Aliaksei Kachanovich,
Kirill Melnikov, 
Go Mishima,
Ulrich Nierste,
Dean J. Robinson,
Yukinari Sumino,
Christopher Wever, and 
Kei Yamamoto 
for helpful discussions.
This work has been supported in part by BMBF under Contract No.~05H15VKKB1.

%%%%%%%%%%%%%%%%%%%%%%%%%%%%%%%%%%%%%%%%%%%%%
% References
%%%%%%%%%%%%%%%%%%%%%%%%%%%%%%%%%%%%%%%%%%%%%
\providecommand{\href}[2]{#2}
\begingroup\raggedright

%%%%%%%%%%%%%%%%%%%%%%%%%%%%%%%%%%%%%%%%%%%%%

\begin{thebibliography}{99}

%\cite{Golob}
\bibitem{Golob}
  B. Golob, K. Trabelsi, P. Urquijo, 
  \href{https://confluence.desy.de/download/attachments/34042032/belle2-note-0021.pdf}{BELLE2-NOTE0021}

%\cite{Aoki:2016frl}
\bibitem{Aoki:2016frl} 
  S.~Aoki {\it et al.},
  %``Review of lattice results concerning low-energy particle physics,''
  \href{http://dx.doi.org/10.1140/epjc/s10052-016-4509-7}{Eur.\ Phys.\ J.\ C {\bf 77}, no. 2, 112 (2017)}
  [arXiv:1607.00299 [hep-lat]].
  %%CITATION = doi:10.1140/epjc/s10052-016-4509-7;%%
  
%\cite{Sirlin:1981ie}
\bibitem{Sirlin:1981ie} 
  A.~Sirlin,
  %``Large m(W), m(Z) Behavior of the O(alpha) Corrections to Semileptonic Processes Mediated by W,''
  \href{http://dx.doi.org/10.1016/0550-3213(82)90303-0}{Nucl.\ Phys.\ B {\bf 196}, 83 (1982).}
  %%CITATION = doi:10.1016/0550-3213(82)90303-0;%%  
  
%\cite{Atwood:1989em}
\bibitem{Atwood:1989em} 
  D.~Atwood and W.~J.~Marciano,
  %``Radiative Corrections and Semileptonic $B$ Decays,''
  \href{http://dx.doi.org/10.1103/PhysRevD.41.1736}{Phys.\ Rev.\ D {\bf 41}, 1736 (1990).}
  %%CITATION = doi:10.1103/PhysRevD.41.1736;%%

%\cite{Becirevic:2009fy}
\bibitem{Becirevic:2009fy} 
  D.~Becirevic and N.~Kosnik,
  %``Soft photons in semileptonic B ---> D decays,''
  \href{http://www.actaphys.uj.edu.pl/sup3/abs/s3p0207.htm?series=sup&vol=3&page=207}{Acta Phys.\ Polon.\ Supp.\  {\bf 3}, 207 (2010)}
  [arXiv:0910.5031 [hep-ph]].
  %%CITATION = ARXIV:0910.5031;%%
  
%\cite{Bernlochner:2010yd}
\bibitem{Bernlochner:2010yd} 
  F.~U.~Bernlochner and H.~Lacker,
  %``A phenomenological model for radiative corrections in exclusive semileptonic B-meson decays to (pseudo)scalar final state mesons,''
  arXiv:1003.1620 [hep-ph].
  %%CITATION = ARXIV:1003.1620;%%
  
%\cite{Bernlochner:2010fc}
\bibitem{Bernlochner:2010fc} 
  F.~U.~Bernlochner and M.~Schonherr,
  %``Comparing different ansatzes to describe electroweak radiative corrections to exclusive semileptonic B meson decays into (pseudo)scalar final state mesons using Monte-Carlo techniques,''
  arXiv:1010.5997 [hep-ph].
  %%CITATION = ARXIV:1010.5997;%%
  
%\cite{Lattice:2015rga}
\bibitem{Lattice:2015rga} 
  J.~A.~Bailey {\it et al.} [MILC Collaboration],
  %``B→D\UTF{2113}ν form factors at nonzero recoil and |V$_{cb}$| from 2+1-flavor lattice QCD,''
  \href{http://dx.doi.org/10.1103/PhysRevD.92.034506}{Phys.\ Rev.\ D {\bf 92}, no. 3, 034506 (2015)}
  [arXiv:1503.07237 [hep-lat]].
  %%CITATION = doi:10.1103/PhysRevD.92.034506;%%

%\cite{Na:2015kha}
\bibitem{Na:2015kha} 
  H.~Na {\it et al.} [HPQCD Collaboration],
  %``$B \rightarrow D l \nu$ form factors at nonzero recoil and extraction of $|V_{cb}|$,''
   \href{http://dx.doi.org/10.1103/PhysRevD.92.054510}{Phys.\ Rev.\ D {\bf 92}, no. 5, 054510 (2015)}
  Erratum: [\href{http://dx.doi.org/10.1103/PhysRevD.93.119906}{Phys.\ Rev.\ D {\bf 93}, no. 11, 119906 (2016)}]
  [arXiv:1505.03925 [hep-lat]].
  %%CITATION = doi:10.1103/PhysRevD.93.119906, 10.1103/PhysRevD.92.054510;%%

%\cite{Kamenik:2008tj}
\bibitem{Kamenik:2008tj}
  J.~F.~Kamenik and F.~Mescia,
  %``$B \to  D \tau \nu$ Branching Ratios: Opportunity for Lattice QCD and Hadron Colliders,''
  \href{http://dx.doi.org/10.1103/PhysRevD.78.014003}{Phys.\ Rev.\ D {\bf 78} (2008) 014003}
  [arXiv:0802.3790 [hep-ph]].
  %%CITATION = doi:10.1103/PhysRevD.78.014003;%%
  
%\cite{Becirevic:2012jf}
\bibitem{Becirevic:2012jf}
  D.~Be\v{c}irevi\'c, N.~Ko\v{s}nik and A.~Tayduganov,
  %``$\bar B\to D\tau\bar \nu_\tau$ vs. $\bar B\to D\mu\bar \nu_\mu$,''
  \href{http://dx.doi.org/10.1016/j.physletb.2012.08.016}{Phys.\ Lett.\ B {\bf 716} (2012) 208}
  [arXiv:1206.4977 [hep-ph]].
  %%CITATION = doi:10.1016/j.physletb.2012.08.016;%%

%\cite{Bigi:2016mdz}
\bibitem{Bigi:2016mdz} 
  D.~Bigi and P.~Gambino,
  %``Revisiting $B\to D \ell \nu$,''
 \href{http://dx.doi.org/10.1103/PhysRevD.94.094008}{Phys.\ Rev.\ D {\bf 94}, no. 9, 094008 (2016)}
  [arXiv:1606.08030 [hep-ph]].
  %%CITATION = doi:10.1103/PhysRevD.94.094008;%%

%\cite{Bernlochner:2017jka}
\bibitem{Bernlochner:2017jka} 
  F.~U.~Bernlochner, Z.~Ligeti, M.~Papucci and D.~J.~Robinson,
  %``Combined analysis of semileptonic $B$ decays to $D$ and $D^*$: $R(D^{(*)})$, $|V_{cb}|$, and new physics,''
 \href{http://dx.doi.org/10.1103/PhysRevD.95.115008}{Phys.\ Rev.\ D {\bf 95}, no. 11, 115008 (2017)}
  [arXiv:1703.05330 [hep-ph]].
  %%CITATION = doi:10.1103/PhysRevD.95.115008;%%
  
%\cite{Jaiswal:2017rve}
\bibitem{Jaiswal:2017rve}
  S.~Jaiswal, S.~Nandi and S.~K.~Patra,
  %``Extraction of $|V_{cb}|$ from $B\to D^{(*)}\ell\nu_\ell$ and the Standard Model predictions of $R(D^{(*)})$,''
  \href{http://dx.doi.org/10.1007/JHEP12(2017)060}{JHEP {\bf 1712} (2017) 060}
  [arXiv:1707.09977 [hep-ph]].
  %%CITATION = doi:10.1007/JHEP12(2017)060;%%

%\cite{Amhis:2016xyh}
\bibitem{Amhis:2016xyh} 
  Y.~Amhis {\it et al.} [HFLAV Collaboration],
  %``Averages of $b$-hadron, $c$-hadron, and $\tau$-lepton properties as of summer 2016,''
 \href{http://dx.doi.org/10.1140/epjc/s10052-017-5058-4}{Eur.\ Phys.\ J.\ C {\bf 77}, no. 12, 895 (2017)}
  [arXiv:1612.07233 [hep-ex]].
  %%CITATION = doi:10.1140/epjc/s10052-017-5058-4;%%

%\cite{Lees:2012xj}
\bibitem{Lees:2012xj} 
  J.~P.~Lees {\it et al.} [BaBar Collaboration],
  %``Evidence for an excess of $\bar{B} \to D^{(*)} \tau^-\bar{\nu}_\tau$ decays,''
  \href{http://dx.doi.org/10.1103/PhysRevLett.109.101802}{Phys.\ Rev.\ Lett.\  {\bf 109}, 101802 (2012)}
  [arXiv:1205.5442 [hep-ex]].
  %%CITATION = doi:10.1103/PhysRevLett.109.101802;%%

%\cite{Lees:2013uzd}
\bibitem{Lees:2013uzd} 
  J.~P.~Lees {\it et al.} [BaBar Collaboration],
  %``Measurement of an Excess of $\bar{B} \to D^{(*)}\tau^- \bar{\nu}_\tau$ Decays and Implications for Charged Higgs Bosons,''
  \href{http://dx.doi.org/10.1103/PhysRevD.88.072012}{Phys.\ Rev.\ D {\bf 88}, no. 7, 072012 (2013)}
  [arXiv:1303.0571 [hep-ex]].
  %%CITATION = doi:10.1103/PhysRevD.88.072012;%%

%\cite{Huschle:2015rga}
\bibitem{Huschle:2015rga} 
  M.~Huschle {\it et al.} [Belle Collaboration],
  %``Measurement of the branching ratio of $\bar{B} \to D^{(\ast)} \tau^- \bar{\nu}_\tau$ relative to $\bar{B} \to D^{(\ast)} \ell^- \bar{\nu}_\ell$ decays with hadronic tagging at Belle,''
   \href{http://dx.doi.org/10.1103/PhysRevD.92.072014}{Phys.\ Rev.\ D {\bf 92}, no. 7, 072014 (2015)}
  [arXiv:1507.03233 [hep-ex]].
  %%CITATION = doi:10.1103/PhysRevD.92.072014;%%
  
\bibitem{Belle1}
  F.~Bernlochner and P.~Goldenzweig, private communication.

\bibitem{Belle2}
  S.~Hirose, private communication.
    
%\cite{Barberio:1990ms}
\bibitem{Barberio:1990ms} 
  E.~Barberio, B.~van Eijk and Z.~Was,
  %``PHOTOS: A Universal Monte Carlo for QED radiative corrections in decays,''
  \href{http://dx.doi.org/10.1016/0010-4655(91)90012-A}{Comput.\ Phys.\ Commun.\  {\bf 66}, 115 (1991).}
  %doi:10.1016/0010-4655(91)90012-A
  %%CITATION = doi:10.1016/0010-4655(91)90012-A;%%
  
%\cite{Barberio:1993qi}
\bibitem{Barberio:1993qi} 
  E.~Barberio and Z.~Was,
  %``PHOTOS: A Universal Monte Carlo for QED radiative corrections. Version 2.0,''
  \href{http://dx.doi.org/10.1016/0010-4655(94)90074-4}{Comput.\ Phys.\ Commun.\  {\bf 79}, 291 (1994).}
  %%CITATION = doi:10.1016/0010-4655(94)90074-4;%%
 
%\cite{Davidson:2010ew}
\bibitem{Davidson:2010ew} 
  N.~Davidson, T.~Przedzinski and Z.~Was,
  %``PHOTOS interface in C++: Technical and Physics Documentation,''
  \href{http://dx.doi.org/10.1016/j.cpc.2015.09.013}{Comput.\ Phys.\ Commun.\  {\bf 199}, 86 (2016)}
  [arXiv:1011.0937 [hep-ph]].
  %%CITATION = doi:10.1016/j.cpc.2015.09.013;%%

\bibitem{footnote1}
  Version 2.13 PHOTOS \cite{Golonka:2005pn} accounts for interferencies of final-state emissions in semileptonic decays.

%\cite{Golonka:2005pn}
\bibitem{Golonka:2005pn} 
  P.~Golonka and Z.~Was,
  %``PHOTOS Monte Carlo: A Precision tool for QED corrections in $Z$ and $W$ decays,''
  \href{http://dx.doi.org/10.1140/epjc/s2005-02396-4}{Eur.\ Phys.\ J.\ C {\bf 45}, 97 (2006)}
  [hep-ph/0506026].
  %%CITATION = doi:10.1140/epjc/s2005-02396-4;%%  
  
%\cite{Cirigliano:2005ms}
\bibitem{Cirigliano:2005ms} 
  V.~Cirigliano and D.~Pirjol,
  %``Factorization in exclusive semileptonic radiative B decays,''
  \href{http://dx.doi.org/10.1103/PhysRevD.72.094021}{Phys.\ Rev.\ D {\bf 72}, 094021 (2005)}
  [hep-ph/0508095].
  %%CITATION = doi:10.1103/PhysRevD.72.094021;%%
  
%\cite{Becirevic:2009aq}
\bibitem{Becirevic:2009aq} 
  D.~Becirevic, B.~Haas and E.~Kou,
  %``Soft Photon Problem in Leptonic B-decays,''
  \href{http://dx.doi.org/10.1016/j.physletb.2009.10.017}{Phys.\ Lett.\ B {\bf 681}, 257 (2009)}
  %doi:10.1016/j.physletb.2009.10.017
  [arXiv:0907.1845 [hep-ph]].
  %%CITATION = doi:10.1016/j.physletb.2009.10.017;%%
  
    
%\cite{Tostado:2015tna}
\bibitem{Tostado:2015tna} 
  S.~L.~Tostado and G.~L\'opez Castro,
  %``Radiative corrections of $O(\alpha )$ to $B^{-} \rightarrow V^{0} \ell ^{-} \bar{\nu }_{\ell }$ decays,''
 \href{http://dx.doi.org/10.1140/epjc/s10052-016-4329-9}{Eur.\ Phys.\ J.\ C {\bf 76}, no. 9, 495 (2016)}
  [arXiv:1510.08020 [hep-ph]].
  %%CITATION = doi:10.1140/epjc/s10052-016-4329-9;%%
 
%\cite{Buras:2012ru}
\bibitem{Buras:2012ru} 
  A.~J.~Buras, J.~Girrbach, D.~Guadagnoli and G.~Isidori,
  %``On the Standard Model prediction for BR(B{s,d} to mu+ mu-),''
 \href{http://dx.doi.org/10.1140/epjc/s10052-012-2172-1}{Eur.\ Phys.\ J.\ C {\bf 72}, 2172 (2012)}
  [arXiv:1208.0934 [hep-ph]].
  %%CITATION = doi:10.1140/epjc/s10052-012-2172-1;%%
  
%\cite{Bordone:2016gaq}
\bibitem{Bordone:2016gaq} 
  M.~Bordone, G.~Isidori and A.~Pattori,
  %``On the Standard Model predictions for $R_K$ and $R_{K^*}$,''
 \href{http://dx.doi.org/10.1140/epjc/s10052-016-4274-7}{Eur.\ Phys.\ J.\ C {\bf 76}, no. 8, 440 (2016)}
  [arXiv:1605.07633 [hep-ph]].
  %%CITATION = doi:10.1140/epjc/s10052-016-4274-7;%%
  
%\cite{Beneke:2017vpq}
\bibitem{Beneke:2017vpq} 
  M.~Beneke, C.~Bobeth and R.~Szafron,
  %``Enhanced electromagnetic correction to the rare $B$-meson decay $B_{s,d} \to \mu^+ \mu^-$,''
  \href{http://dx.doi.org/10.1103/PhysRevLett.120.011801}{Phys.\ Rev.\ Lett.\  {\bf 120}, no. 1, 011801 (2018)}
  [arXiv:1708.09152 [hep-ph]].
  %%CITATION = doi:10.1103/PhysRevLett.120.011801;%%
  
%\cite{Patrignani:2016xqp}
\bibitem{Patrignani:2016xqp} 
  C.~Patrignani {\it et al.} [Particle Data Group],
  %``Review of Particle Physics,''
  \href{http://dx.doi.org/10.1088/1674-1137/40/10/100001}{Chin.\ Phys.\ C {\bf 40}, no. 10, 100001 (2016).}
  %%CITATION = doi:10.1088/1674-1137/40/10/100001;%%
     
%\cite{Isidori:2007zt}
\bibitem{Isidori:2007zt} 
  G.~Isidori,
  %``Soft-photon corrections in multi-body meson decays,''
 \href{http://dx.doi.org/10.1140/epjc/s10052-007-0487-0}{Eur.\ Phys.\ J.\ C {\bf 53}, 567 (2008)}
  [arXiv:0709.2439 [hep-ph]].
  %%CITATION = doi:10.1140/epjc/s10052-007-0487-0;%%
   
%\cite{Jauch}
\bibitem{Jauch} 
  J. M.~Jauch and  F.~Rohrlich,
  %``The infrared divergence,''
 \href{http://doi.org/10.5169/seals-112533}{Helv. Phys. Acta\ {\bf 27}, 613 (1954).}
  
%\cite{Yennie:1961ad}
\bibitem{Yennie:1961ad} 
  D.~R.~Yennie, S.~C.~Frautschi and H.~Suura,
  %``The infrared divergence phenomena and high-energy processes,''
  \href{http://dx.doi.org/10.1016/0003-4916(61)90151-8}{Annals Phys.\  {\bf 13}, 379 (1961).}
  %%CITATION = doi:10.1016/0003-4916(61)90151-8;%%

%\cite{Weinberg:1965nx}
\bibitem{Weinberg:1965nx} 
  S.~Weinberg,
  %``Infrared photons and gravitons,''
   \href{http://dx.doi.org/10.1103/PhysRev.140.B516}{Phys.\ Rev.\  {\bf 140}, B516 (1965).}
  %%CITATION = doi:10.1103/PhysRev.140.B516;%%
  
%\cite{Sommerfeld}
\bibitem{Sommerfeld}
  A.~Sommerfeld, 
  Annalen der Physik\ {\bf 403}, 257 (1931).
  
%\cite{Hryczuk:2011tq}
\bibitem{Hryczuk:2011tq} 
  A.~Hryczuk,
  %``The Sommerfeld enhancement for scalar particles and application to sfermion co-annihilation regions,''
  \href{http://dx.doi.org/10.1016/j.physletb.2011.04.016}{Phys.\ Lett.\ B {\bf 699}, 271 (2011)}
  [arXiv:1102.4295 [hep-ph]].
  %%CITATION = doi:10.1016/j.physletb.2011.04.016;%%
  
%\cite{Falgari:2012sq}
\bibitem{Falgari:2012sq} 
  P.~Falgari, C.~Schwinn and C.~Wever,
  %``Finite-width effects on threshold corrections to squark and gluino production,''
  \href{http://dx.doi.org/10.1007/JHEP01(2013)085}{JHEP {\bf 1301}, 085 (2013)}
  [arXiv:1211.3408 [hep-ph]].
  %%CITATION = doi:10.1007/JHEP01(2013)085;%%
  
  %\cite{Kubis:2006nh}
\bibitem{Kubis:2006nh} 
  B.~Kubis, E.~H.~Muller, J.~Gasser and M.~Schmid,
  %``Aspects of radiative K+(e3) decays,''
  \href{http://dx.doi.org/10.1140/epjc/s10052-007-0215-9}{Eur.\ Phys.\ J.\ C {\bf 50}, 557 (2007)}
  [hep-ph/0611366].
  %%CITATION = doi:10.1140/epjc/s10052-007-0215-9;%%
  
%\cite{Beenakker:1988jr}
\bibitem{Beenakker:1988jr} 
  W.~Beenakker and A.~Denner,
  %``Infrared Divergent Scalar Box Integrals With Applications in the Electroweak Standard Model,''
  \href{http://dx.doi.org/10.1016/0550-3213(90)90636-R}{Nucl.\ Phys.\ B {\bf 338}, 349 (1990).}
  %%CITATION = doi:10.1016/0550-3213(90)90636-R;%%
  
%\cite{Hahn:1998yk}
\bibitem{Hahn:1998yk} 
  T.~Hahn and M.~Perez-Victoria,
  %``Automatized one loop calculations in four-dimensions and D-dimensions,''
  \href{http://dx.doi.org/10.1016/S0010-4655(98)00173-8}{Comput.\ Phys.\ Commun.\  {\bf 118}, 153 (1999)}
  [hep-ph/9807565].
  %%CITATION = doi:10.1016/S0010-4655(98)00173-8;%%
  
%\cite{Patel:2015tea}
\bibitem{Patel:2015tea} 
  H.~H.~Patel,
  %``Package-X: A Mathematica package for the analytic calculation of one-loop integrals,''
  \href{http://dx.doi.org/10.1016/j.cpc.2015.08.017}{Comput.\ Phys.\ Commun.\  {\bf 197}, 276 (2015)}
  [arXiv:1503.01469 [hep-ph]].
  %%CITATION = doi:10.1016/j.cpc.2015.08.017;%%
  
%\cite{Kubis:2010mp}
\bibitem{Kubis:2010mp} 
  B.~Kubis and R.~Schmidt,
  %``Radiative corrections in $K -> \pi \l^+ \l^-$ decays,''
  \href{http://dx.doi.org/10.1140/epjc/s10052-010-1442-z}{Eur.\ Phys.\ J.\ C {\bf 70}, 219 (2010)}
  [arXiv:1007.1887 [hep-ph]].
  %%CITATION = doi:10.1140/epjc/s10052-010-1442-z;%%
  
\end{thebibliography}
\end{document}